\documentclass[12pt]{article}
\usepackage{amsmath}
\usepackage{graphicx}
\usepackage{enumerate}
\usepackage{natbib}
\usepackage{multibib}
\usepackage{silence}
\usepackage{bm}
\usepackage{bbm}
\usepackage{url} % not crucial - just used below for the URL 
\usepackage[colorlinks, citecolor=blue]{hyperref}
\usepackage{amsthm}
\usepackage{amsfonts}
\usepackage{amssymb}
\usepackage{multirow}
\usepackage{booktabs}
\usepackage[justification=centering]{caption}
\usepackage{threeparttable}
\usepackage{graphicx}
\usepackage{float}
\usepackage{comment}
\usepackage{subcaption}
\usepackage{cleveref}
\usepackage{threeparttable}
\usepackage{xcolor}
\usepackage{colortbl}
\usepackage{xr}
\newtheorem{assumption}{Assumption}
\newtheorem{theorem}{Theorem}
\newtheorem{proposition}{Proposition}

\theoremstyle{definition}
\newtheorem{example}{Example}
\crefname{assumption}{Assumption}{Assumptions}
\Crefname{assumption}{Assumption}{Assumptions}

\newcommand{\blind}{1}

\newcommand{\Y}[3]{Y_{#1 \rightarrow #2}(#3)}
\newcommand{\Yo}[2]{Y_{#1 \rightarrow #2}^{\mathrm{obs}}}
\newcommand{\z}{\bm \zeta}
\newcommand{\pr}{\mathrm{pr}}
\newcommand{\sen}{\mathrm{sen}}
\newcommand{\rec}{\mathrm{rec}}
\newcommand{\var}{\mathrm{var}}

\newcommand{\argmin}{\mathop{\arg \min}}
\newcommand{\ora}{\mathrm{ora}}
\newcommand{\adj}{\mathrm{adj}}

\newcommand{\e}[3]{e_{#1\rightarrow #2}(#3)}
\newcommand{\ehat}[3]{\hat e_{#1\rightarrow #2}(#3)}
\newcommand{\varu}[1]{\mathrm{var}^{U #1}}
\newcommand{\varhatu}[1]{\widehat{\mathrm{var}}^{U #1}}
\newcommand{\sig}[1][]{\sigma_{n #1}}
\newcommand{\sigora}[1][]{\sigma_{n #1,\ora}}

\begin{document}

\def\spacingset#1{\renewcommand{\baselinestretch}%
{#1}\small\normalsize} \spacingset{1}

%%%%%%%%%%%%%%%%%%%%%%%%%%%%%%%%%%%%%%%%%%%%%%%%%%%%%%%%%%%%%%%%%%%%%%%%%%%%%%

\if1\blind
{
  \title{\bf Design-based edge-level causal inference with machine learning assisted covariate adjustment}
  \author{
Haoyang Yu$^{1}$,
Yilin Li$^{2}$,
Lu Deng$^{2}$,
Yong Wang$^{2}$,
Xin Lu$^{3}$\footnote{Corresponding author. Email: \texttt{xinl@wustl.edu}.},\\
and Hanzhong Liu$^{1}$\footnote{Co-corresponding author. Email: \texttt{lhz2016@tsinghua.edu.cn}. Hanzhong Liu was supported by the Beijing Natural Science Foundation (F251001)
and the National Natural Science Foundation of China (12531012).}
\\[0.5cm]
$^{1}$ Department of Statistics and Data Science, Tsinghua University \\
$^{2}$ Tencent Inc. \\
$^{3}$ Department of Statistics and Data Science,\\ Washington University in St. Louis
}
    \date{}
\renewcommand{\thefootnote}{\fnsymbol{footnote}}
\setcounter{footnote}{0}
  \maketitle
} \fi

\if0\blind
{
  \bigskip
  \bigskip
  \bigskip
  \begin{center}
    {\LARGE\bf Design-based edge-level causal inference with machine learning assisted covariate adjustment}
\end{center}
  \medskip
} \fi

%\bigskip
\begin{abstract}

We study design-based causal inference for edge-level outcomes in directed networks under dyadic interference. In this setting, outcomes are defined on directed edges and depend on the joint treatment assignments of pairs of units, inducing a complex dependence structure that invalidates standard estimation and inference procedures developed for node-level data. We construct Horvitz--Thompson estimators for a general class of edge-level causal effects and establish their asymptotic normality under mild regularity conditions. To enable valid inference, we develop variance estimators that exploit identifiable components of network dependence, yielding substantially less conservative bounds than classical approaches. To improve efficiency, we incorporate auxiliary covariates through a sample splitting and cross-fitting procedure. A key technical challenge is that standard two-fold sample splitting fails in the presence of edge-level outcomes due to the dependence induced by shared units. To address this issue, we introduce a three-fold sample splitting and cross-fitting scheme that restores the conditional independence required for unbiased estimation. Under a stability condition, the resulting covariate-adjusted estimator is asymptotically normal and accommodates both linear adjustment and flexible machine learning methods. We further introduce a calibration step that guarantees no asymptotic efficiency loss relative to the unadjusted estimator. Simulation studies and a real-data application confirm the theoretical results and demonstrate substantial efficiency gains.
\end{abstract}

\noindent%
\vfill

\newpage
\spacingset{1.9} % DON'T change the spacing!

\section{Introduction}
\label{sec:intro}

Causal inference provides rigorous tools for evaluating intervention effects in areas such as clinical trials, healthcare, economics, and public policy. Central to modern causal inference is the potential outcomes framework \citep{Neyman1923, rubin1974}, under which each unit is associated with a collection of potential outcomes corresponding to different treatment assignments. A standard assumption is the Stable Unit Treatment Value Assumption (SUTVA) \citep{Rubin1980}, which consists of two components: (1) no interference between units, and (2) no hidden variations in treatment; thus, each treatment level corresponds to a unique potential outcome. However, in modern applications, experimental units are often interdependent through networks, such as social networks formed through family, community, or friendship ties. In such settings, SUTVA may fail to hold, as interference is often inherent to the underlying system structure.

To balance tractability and realism, researchers have proposed various relaxations of SUTVA that incorporate network structure, including partial interference that restricts interference to predefined groups \citep{hudgens2008,tchetgen2012,liu2014}, exposure mapping approaches in which outcomes depend on summary statistics of neighboring treatments \citep{aronow2017estimating,leung2020treatment,savje2021}, and low-order interaction models that allow for unit-specific interaction effects \citep{sussman2017elements,yu2022estimating,cortez2023exploiting}.

Most existing frameworks are formulated for node-level outcomes. In many applications, however, the primary outcomes of interest are defined on edges, representing interactions or relationships between pairs of units. Examples include communication events in email networks \citep{diesner2005communication}, international trade flows \citep{carlson2024dyadic}, and patterns of disease transmission in epidemiological networks \citep{balcan2009multiscale}. The prevalence of such outcomes has motivated recent methodological developments for dyadic data, including methods for association testing, interference detection, causal inference, and ordered dyadic outcomes \citep{shi2024asymptotic,li2025causal,muris2025}. 

From a causal inference perspective, edge-level outcomes introduce challenges beyond those encountered in node-level settings. Each outcome depends on the joint treatment assignments of both the sender and the receiver, and edge observations sharing common endpoints exhibit a distinctive dependence structure. Consequently, methods developed for node-level outcomes do not directly extend to directed-edge settings, and design-based causal inference for edge-level outcomes remains largely underexplored.

\textbf{Our contributions:} To address these challenges, we develop a design-based covariate-adjusted causal inference framework for edge-level outcomes in directed networks, treating the potential outcomes and network structure as fixed while regarding the treatment assignment as the sole source of randomness. Our contributions are fourfold.

First, we define a general class of edge-level causal effects and establish convergence rates and central limit theorems for the corresponding Horvitz--Thompson estimators. Our assumptions regarding the network are weaker than those in the existing literature. Second, we develop variance estimators that avoid the over-conservativeness induced by classical Cauchy--Schwarz arguments by exploiting identifiable components of the edge-level dependence structure. Third, we propose a machine learning-assisted augmented inverse probability weighting (AIPW)-type estimator that incorporates edge-level covariates to improve efficiency. Because edge-level outcomes sharing common endpoints are dependent, standard two-fold sample splitting arguments developed for node-level outcomes are no longer valid. To overcome this issue, we introduce a tailored three-fold sample splitting and cross-fitting procedure that restores the conditional independence required for unbiased estimation, even under misspecified prediction models. Under a stability condition on the fitted models, we establish the asymptotic normality of the resulting covariate-adjusted estimator and verify the condition for both linear and nonparametric regression methods. Our theory accommodates high-dimensional covariates. In particular, when the degree of each unit in the network is uniformly bounded, the covariate dimension \(d\) is allowed to diverge at the rate \(o(n)\), where \(n\) denotes the sample size. Finally, we introduce a calibration step that guarantees no efficiency loss relative to the unadjusted estimator. Simulation studies and a real-data application demonstrate the validity of the proposed inference procedures and the substantial efficiency gains achieved by covariate adjustment.

\textbf{Related literature:} Our work is grounded in the design-based framework for randomized experiments, in which potential outcomes are treated as fixed and randomization serves as the sole source of uncertainty \citep{Neyman1923,rubin1974,imbens2015causal,ding2024first}. A key advantage of this framework is that valid inference follows directly from the randomization design and does not rely on correctly specifying either a covariate model or an outcome model. Recent work has begun to investigate design-based causal inference for edge-level outcomes under dyadic interference \citep{deng2024unbiased,li2025causal}. Building on this line of research, we derive sharper estimable variance bounds, incorporate covariate adjustment to improve efficiency, and establish asymptotic theory using a new proof technique developed by \citet{koike2023high}, which substantially relaxes the required network sparsity conditions.

Our work is also closely related to the literature on covariate adjustment and model-assisted inference in randomized experiments \citep{freedman2008regression,lin2013agnostic,bloniarz2016lasso,liu2020regression,guo2023generalized}, where working models are used to improve precision while still allowing model misspecification. More recently, sample splitting and cross-fitting methods have been developed to preserve unbiased treatment effect estimation while incorporating flexible machine learning predictions to improve efficiency for node-level outcomes \citep{su2023decorrelation,lu2025conditional}. These arguments, however, do not directly extend to edge-level outcomes because edge-level dyadic observations sharing common endpoints are inherently dependent. This challenge motivates our three-fold sample splitting scheme and the accompanying stability analysis. Taken together, these connections and distinctions position our framework as a design-based approach for efficient and robust inference on edge-level causal effects with flexible machine learning-assisted covariate adjustment in randomized experiments under dyadic interference.

The remainder of the paper is organized as follows. \Cref{sec:setting} introduces the framework and notation. \Cref{sec:estimation} presents Horvitz--Thompson estimators for the causal effects. \Cref{sec:inference} establishes the asymptotic normality of the Horvitz--Thompson estimators and develops variance estimators. \Cref{sec:covariate} introduces covariate adjustment methods to improve efficiency. The performance of the proposed methods is evaluated through simulation studies in \Cref{sec:simulation} and a real-data application in \Cref{sec:realdata}. \Cref{sec:discussion} concludes with a discussion. All proofs are deferred to the Supplementary Material.

\section{Framework and notation}
\label{sec:setting}

\subsection{Edge-level potential outcomes and causal estimands}

Consider a randomized experiment on $n$ units connected through a directed network with adjacency matrix $\bm{A} = (A_{ij})$, where $A_{ii}=0$ and $A_{ij}\in \{0,1\}$ indicates the presence of a directed edge from unit $i$ to unit $j$. Each unit $i$ is independently assigned a binary treatment. Specifically, let $Z_i \stackrel{\textnormal{i.i.d.}}{\sim} \textnormal{Bernoulli}(r)$ denote the treatment indicator for unit $i$, where $r\in(0,1)$, and let $\bm{Z}=(Z_1,\dots,Z_n)^\top$ denote the treatment assignment vector for all units. Throughout the paper, we adopt the design-based framework, under which the potential outcomes are treated as fixed and all randomness arises solely from the treatment assignment vector $\bm{Z}$.

The potential outcome for edge $i \to j$, denoted by $\Y{i}{j}{\bm z}$, generally depends on the entire treatment assignment vector $\bm z$. To reduce the dimensionality of the potential outcomes, we impose the \textbf{dyadic interference} assumption, under which the outcome on edge $i \to j$ depends only on the treatment assignments of its sender $i$ and receiver $j$. Under this assumption, the potential outcome can be written as $\Y{i}{j}{z_i, z_j}$, and, by the consistency assumption, the observed outcome is given by $\Yo{i}{j} = \Y{i}{j}{Z_i, Z_j}$.

Although dyadic interference is a strong assumption, it has been adopted in the literature \citep{deng2024unbiased,li2025causal} and provides a natural starting point for rigorous edge-level causal inference. Similar forms of dependence also arise frequently in recent work on network and dyadic data analysis \citep{li2022random,shi2024asymptotic,muris2025}, where edge formation or edge-level outcomes are often modeled as functions of the characteristics of the two endpoints.

When edge-level outcomes are aggregated to the node level, the dyadic interference assumption is weaker than the Heterogeneous Additive Treatment Effect Model (HATEM) widely used in the network interference literature \citep{sussman2017elements,yu2022estimating,lu2024estimation}. Compared with HATEM, the edge-level potential outcomes framework under dyadic interference offers two important advantages. First, it accommodates a richer class of causal estimands through different weighting vectors $\bm \omega$. Second, it can achieve substantially greater statistical efficiency; see Section~\ref{sec:compare-hatem} for a detailed comparison. Extensions to more general interference structures via exposure mappings are provided in \Cref{sec.sm:exposure} of the Supplementary Material.

We define the adjacency matrix $\bm A$ based on whether information may potentially be shared or transmitted between two units. Specifically, $A_{ij}=1$ if $\Y{i}{j}{z_i, z_j} \neq 0$ for at least one $(z_i, z_j)\in \{0,1\}^2$, indicating the presence of information sharing or transmission from unit $i$ to unit $j$, and $A_{ij}=0$ otherwise. Importantly, the underlying adjacency matrix $\bm A$ itself may be unobserved.

We assume $\Y{i}{i}{z_i,z_i}=0$ and allow for asymmetry, in the sense that $\Y{i}{j}{z_i, z_j}$ need not equal $\Y{j}{i}{z_j, z_i}$. Consequently, each directed edge has four potential outcomes corresponding to the four possible joint treatment assignments: $\Y{i}{j}{1,1}$, $\Y{i}{j}{1,0}$, $\Y{i}{j}{0,1}$, and $\Y{i}{j}{0,0}$. This framework naturally arises in many social network settings, which we illustrate through the following three examples.

\begin{example}
There is a growing interest in developing recommendation algorithms that encourage active sharing within social networks to increase user engagement. 
To assess the causal effect of these algorithms, large social media platforms like WeChat conduct large-scale experiments in which an updated recommendation algorithm (treatment) is randomly assigned to users (units) \citep{li2025causal}. 
Here, unit $i$ is a user with treatment $Z_i$ indicating the assigned algorithm, and the edge-level outcome $Y_{i\to j}$ measures sender-to-receiver interaction intensity, including sharing duration, sharing impressions, or successful sharing events from $i$ to $j$.
The dyadic interference assumption states that this outcome depends only on the treatments of the sender and receiver, and is not affected by the treatment assignments of other units.
\end{example}

\begin{example}
Organizational economics often examines the impact of organizational hierarchy on employee collaboration. 
For example, in observational studies using the Enron email corpus \citep{diesner2005communication}, the treatment is employee's role (whether the unit is a manager or subordinate), and the outcome ($Y_{i\to j}$) is the number of emails sent from employee $i$ to employee $j$. 
In this context, the dyadic interference assumption posits that communication between two individuals depends on their roles (e.g., manager-to-subordinate versus peer-to-peer) and is not affected by the roles of other employees in the network.
\end{example}

\begin{example}
In public policy studies, a key goal is to understand the impact of regional policies on inter-regional mobility. 
For example, a geographic region such as a county can be a unit, and a public health directive like a stay-at-home order can be the treatment \citep{holtz2020interdependence}. 
The edge-level outcome $Y_{i \to j}$ can then be defined as the travel volume from region $i$ to region $j$. 
In this case, the dyadic interference assumption posits that travel between two regions is determined solely by the policies of the origin region $i$ and the destination region $j$, and is independent of policies in other regions.
\end{example}

Let $\z=(z_{\sen},z_{\rec}) \in \{0,1\}^2$ denote a generic joint treatment configuration. Define $\theta(\z) = n^{-1}\sum_{(i,j):\, i\neq j}\Y{i}{j}{\z}$ as the average potential outcome under treatment configuration $\z$. We consider the causal estimand $\tau = \sum_{\z} \omega(\z)\theta(\z)$, where the weight vector $\bm \omega = (\omega(1,1), \omega(1,0), \omega(0,1), \omega(0,0))$ satisfies $\sum_{\z}\omega(\z)=0$ and $|\omega(\z)|\leq 1$ for all $\z$. By choosing different weights, $\tau$ can represent a variety of causal effects of interest. For example, the sender effect is $\tau_{\sen}=\theta(1,0)-\theta(0,0)$, corresponding to $\bm\omega=(0,1,0,-1)$; the receiver effect is $\tau_{\rec}=\theta(0,1)-\theta(0,0)$, corresponding to $\bm\omega=(0,0,1,-1)$; and the total effect is $\tau_{\mathrm{tot}}=\theta(1,1)-\theta(0,0)$, corresponding to $\bm\omega=(1,0,0,-1)$.

\subsection{Comparison with traditional node-level settings}
\label{sec:compare-hatem}

In this section, we compare our edge-level framework with the traditional node-level framework under network interference. One natural way to connect the two formulations is to view a node-level outcome as an aggregation of edge-level outcomes: $Y_i(\bm z)=\sum_{j=1}^n \Y{i}{j}{z_i,z_j}$.
Without additional structural restrictions, node-level aggregation may discard the sender-receiver treatment-configuration information needed for unbiased edge-level estimation.
This difficulty is illustrated by \citet[Theorem~1]{li2025causal}, who showed that conventional Horvitz--Thompson type estimators based on unit-level aggregates are generally biased for the global treatment effect under dyadic interference.

One common way to obtain a tractable node-level formulation is to impose the Heterogeneous Additive Treatment Effect Model (HATEM) \citep{sussman2017elements,yu2022estimating}, under which $Y_i(\bm z)=\alpha_i+\theta_i z_i+\sum_{j=1}^n\gamma_{ij}z_j$. Here, $\alpha_i$ denotes the baseline outcome, $\theta_i$ captures the direct effect, and $\gamma_{ij}$ represents the spillover effect from unit $j$ to unit $i$. Under the above aggregation representation, the HATEM parameters can be expressed as
\begin{align*}
    \alpha_i&=\sum_{j=1}^n\Y{i}{j}{0,0}, \quad
    \theta_i=\sum_{j=1}^n\{\Y{i}{j}{1,0}-\Y{i}{j}{0,0}\}, \quad
    \gamma_{ij}=\Y{i}{j}{0,1}-\Y{i}{j}{0,0},
\end{align*}
provided that the sender and receiver treatments do not interact with each edge, namely,
$
\Y{i}{j}{1,1}-\Y{i}{j}{1,0}-\Y{i}{j}{0,1}+\Y{i}{j}{0,0}=0.
$
Therefore, HATEM implicitly rules out endpoint interaction effects in the underlying edge-level potential outcomes.

This restriction can be relaxed by extending the node-level model to include pairwise interaction terms: $Y_i(\bm z)=\alpha_i+\theta_iz_i+\sum_{j=1}^n\gamma_{ij}z_j+\sum_{j=1}^n\delta_{ij}z_iz_j$,
where, under the edge-level aggregation representation, $\delta_{ij}=\Y{i}{j}{1,1}-\Y{i}{j}{1,0}-\Y{i}{j}{0,1}+\Y{i}{j}{0,0}$.
Even after this extension, however, working directly with edge-level potential outcomes retains two advantages.
First, it permits a richer class of causal estimands through different weighting vectors $\bm \omega$, including sender, receiver, and total effects.
Second, and more importantly, it can be more statistically efficient.
The Horvitz--Thompson estimator for the total effect in the edge-level framework has variance of order $O(n\rho_n^2)$, as shown in \Cref{proposition.combination}, where $\rho_n = \sum_{(i,j):i\neq j} A_{ij}/n^2$ denotes the network density. 
Under the same assumptions, the estimator under the HATEM formulation has variance of order $O\{(\max_i m_i^2)\, n\rho_n^2\}$ \citep{lu2024estimation}, where $m_i = \sum_{j=1}^n (A_{ij}+A_{ji})$ is the total degree of unit $i$. This increase arises from degree-dependent rescaling of the HATEM parameters under the edge-level representation: quantities that are originally $O(1)$ (e.g., $\alpha_i$, $\theta_i$) scale to $O(m_i)$, while degree-normalized terms (e.g., $\gamma_{ij}$) scale to $O(1)$.
As a result, the contribution of each parameter is amplified by a factor of $m_i$, leading to an additional $\max_i m_i^2$ factor in the variance.

\section{Estimation}
\label{sec:estimation}

For any treatment configuration $\z=(z_{\sen},z_{\rec})\in\{0,1\}^2$, the Horvitz--Thompson estimator of the average potential outcome $\theta(\z)$ is
\begin{align*}
\hat \theta(\z)=\frac{1}{n}\sum_{(i,j):i\neq j} \frac{I_{ij}(\z)}{\pi(\z)}\Yo{i}{j}
=\frac{1}{n}\sum_{(i,j):i\neq j} \frac{I_{ij}(\z)}{\pi(\z)}\Y{i}{j}{\z},
\end{align*}
where $I_{ij}(\z)$ is the indicator for whether $(Z_i,Z_j)$ equals $\z$ and $\pi(\z)$ is the corresponding probability.
To study its properties, we first decompose $I_{ij}(\z)/\pi(\z) - 1$ as
\begin{align*}
\frac{(2z_{\sen}-1)(Z_i - r_1)}{\pi(z_{\sen})}
+\frac{(2z_{\rec}-1)(Z_j - r_1)}{\pi(z_{\rec})} + \frac{(2z_{\sen}-1)(2z_{\rec}-1)(Z_i - r_1)(Z_j - r_1)}{\pi(z_{\sen})\pi(z_{\rec})},
\end{align*}
where $\pi(z_i)=z_ir_1+(1-z_i)r_0$ with $r_1=r$ and $r_0=1-r$.
This yields the decomposition $\hat \theta(\z)-\theta(\z)=(\hat \theta_1-\theta_1)+(\hat \theta_2-\theta_2)$, where $\hat \theta_1(\z)-\theta_1(\z)=n^{-1}\sum_{i=1}^n (Z_i - r_1)K_i(\z)$ and $\hat \theta_2(\z)-\theta_2(\z)=n^{-1}\sum_{(i,j):i\neq j} (Z_i - r_1)(Z_j - r_1)K_{ij}(\z)$, with
\begin{align*}
K_i(\z)&=\sum_{j=1}^n\left\{\frac{2z_{\sen}-1}{\pi(z_{\sen})}\Y{i}{j}{\z}
+ \frac{2z_{\rec}-1}{\pi(z_{\rec})}\Y{j}{i}{\z}\right\},\\
K_{ij}(\z)& = \frac{(2z_{\sen}-1)(2z_{\rec}-1)}{\pi(z_{\sen})\pi(z_{\rec})}\Y{i}{j}{\z}.
\end{align*}
The plug-in estimator of $\tau$ is $\hat \tau=\sum_{\z}\omega(\z)\hat \theta(\z)$, and the estimation error $\hat \tau-\tau$ can be similarly decomposed as $\hat \tau-\tau=(\hat \tau_1-\tau_1)+(\hat \tau_2-\tau_2)$, where
\begin{align*}
    \hat \tau_1-\tau_1&=\frac 1n\sum_{i=1}^n(Z_i-r_1)K_i,\quad K_i=\sum_{\z}\omega(\z)K_i(\z),\\
    \hat \tau_2-\tau_2&=\frac 1n\sum_{(i,j): i\neq j}(Z_i-r_1)(Z_j-r_1)K_{ij},\quad K_{ij}=\sum_{\z}\omega(\z)K_{ij}(\z).
\end{align*}
Since $E\{(Z_i-r_1)(Z_j-r_1)(Z_k-r_1)\}=0$ whenever $i,j,k$ are not all equal, $\hat \tau_1$ and $\hat \tau_2$ are uncorrelated, and the variance of $\hat \tau$ can be decomposed as $\sig^2 : = \var(\hat \tau)= \sig[,1]^2+\sig[,2]^2$, where $\sig[,1]^2:=\var(\hat{\tau}_1) = n^{-2} r_1r_0\sum_{i=1}^nK_i^2$ and $\sig[,2]^2:=\var(\hat{\tau}_2) = (2n^2)^{-1}r_1^2r_0^2\sum_{(i,j):i\neq j}\left(K_{ij}+K_{ji}\right)^2$.
This decomposition is important for several reasons: it facilitates proving a central limit theorem under weaker assumptions, helps motivate the construction of the variance estimator, and allows for a more straightforward inspection of the variance estimator's order.

To formalize the asymptotic behavior of $\hat \tau$, we first introduce the following assumptions.
Throughout, $C$ denotes a generic constant independent of $n$ whose value may change from line to line. Let $\lambda_{\max}(\bm B)$ denote the maximum eigenvalue of the matrix $\bm B$.

\begin{assumption}
    \label{assumption.bounded}
    The potential outcomes are uniformly bounded: there exists a constant $C$ independent of $n$, such that $|\Y{i}{j}{\z}|\leq C$ for all $1\leq i\neq j\leq n$ and $\z\in\{0,1\}^2$.
\end{assumption}

\begin{assumption}
    \label{assumption.operator}
    The total number of directed edges satisfies $m \ge C n$ for some constant $C>0$ independent of $n$, where $m=\sum_{(i,j):i\neq j}A_{ij}$ is the total number of directed edges,
    and the adjacency matrix satisfies $\|\bm A\|_{2} = O(n\rho_n)$, where 
    $\|\bm A\|_2 = \sqrt{\lambda_{\max}(\bm A^{\top}\bm A)}$ denotes the matrix $2$-norm (operator norm).
\end{assumption}

\Cref{assumption.operator} requires that the network is sufficiently connected ($m \geq Cn$),
and that the operator norm of $\bm A$ grows at most linearly with network size and density. Under \Cref{assumption.operator}, we have $\sum_{i=1}^nm_{i,\mathrm{out}}^2\leq n\|\bm A\|_{2}^2= O(n^3\rho_n^2)$ and $\sum_{i=1}^nm_{i,\mathrm{in}}^2\leq n\|\bm A\|_{2}^2=O(n^3\rho_n^2)$,
where $m_{i,\mathrm{out}}=\sum_{j=1}^nA_{ij}$ and $m_{i,\mathrm{in}}=\sum_{j=1}^nA_{ji}$ are the out-degree and in-degree of unit $i$, respectively. 

\begin{proposition}
    \label{proposition.combination}
    (i) $E(\hat{\tau}) = \tau$. (ii) Under Assumptions \ref{assumption.bounded}--\ref{assumption.operator}, we have $\sig^2=O(n\rho_n^2)$.
\end{proposition}

\Cref{proposition.combination}(i) shows the unbiasedness of $\hat{\tau}$, and \Cref{proposition.combination}(ii) provides the order of the variance. To ensure consistency of $\hat{\tau}$, we require $\rho_n = o(n^{-1/2})$, which aligns with the requirement established in \cite{li2025causal} and is standard in many classical network interference settings \citep{savje2021,li2022random,lu2024estimation}.

\section{Inference}
\label{sec:inference}

In this section, we study statistical inference for $\tau$. 
First, we establish the asymptotic normality of $\hat \tau$ under appropriate regularity conditions. 
Second, we develop variance estimators that leverage the network structure to enable valid inference.
To establish the asymptotic normality of our estimator, we require the following two conditions.

\begin{assumption}
    \label{assumption.lindeberg}
    $\max_i m_{i,\mathrm{in}}^{2}=o(n^{3}\rho_n^{2})$ and $\max_i m_{i,\mathrm{out}}^{2}=o(n^{3}\rho_n^{2})$.   
\end{assumption}

\begin{assumption}
    \label{assumption.sigma}
%    At least one of the following conditions is satisfied:
    $\liminf_{n \to \infty} \sig^2/(n \rho_n^2 ) > 0$.
\end{assumption}

Under \Cref{assumption.operator}, we obtain 
$\sum_{i=1}^n m_{i,\mathrm{in}}^2 = O(n^3 \rho_n^2)$ and 
$\sum_{i=1}^n m_{i,\mathrm{out}}^2 = O(n^3 \rho_n^2)$, 
so \Cref{assumption.lindeberg} functions as a Lindeberg-type condition ensuring asymptotic normality.  
In contrast, \citet[Theorem 4]{li2025causal} imposed a stronger requirement  
$\max_i\{m_{i,\mathrm{in}}^2, m_{i,\mathrm{out}}^2\} = o(n^{7/3}\rho_n^2)$.  
We can use a weaker condition because we leverage the decomposition in \Cref{sec:estimation} together with the alternative approaches of \citet{de1990central} and \citet{koike2023high} to establish the central limit theorem. 
\Cref{assumption.sigma} ensures that the asymptotic variance of the Horvitz--Thompson estimator remains non-degenerate after appropriate rescaling.

\begin{theorem}
    \label{theorem.clt}
    Under Assumptions \ref{assumption.bounded}--\ref{assumption.sigma}, we have $(\hat \tau-\tau)/\sqrt{\var(\hat \tau)} \stackrel{d}{\rightarrow}\mathcal N(0,1)$.
\end{theorem}

Based on \Cref{theorem.clt}, valid confidence intervals require estimation of the unknown variance $\var(\hat \tau)$, which can be expressed as a quadratic form: $\var(\hat \tau) = n^{-2} r_1 r_0 \bm c^{\top} \tilde{\bm Q} \bm c$, where $\tilde{\bm Q}$ is a symmetric matrix defined in \Cref{eq:Q} of the Supplementary Material, and $\bm c$ is a $4n^2 \times 1$ vector with entries $c_{4(i-1)n + 4(j-1) + l} = \omega(\z)\Y{i}{j}{\z}$ with $l=1,2,3,4$ corresponding respectively to $(1,1),(1,0),(0,1),(0,0)$.
A fundamental challenge is that the variance involves cross-products of potential outcomes, such as $Y_{i\to j}(z_1,z_2)Y_{i'\to j'}(z_3,z_4)$. Standard approaches construct conservative variance estimators by replacing $\tilde{\bm Q}$ with a diagonal matrix $\tilde{\bm Q}^d \succeq \tilde{\bm Q}$ using Cauchy--Schwarz inequalities. Here, $\tilde{\bm Q}^d \succeq \tilde{\bm Q}$ means that $\tilde{\bm Q}^d - \tilde{\bm Q}$ is positive semidefinite. Although valid, this strategy is often overly conservative because it treats all cross-products as unobservable, even though some are in fact identifiable from the observed data.

Our main insight is that a cross-product of potential outcomes is identifiable if and only if it does not impose conflicting treatment assignments on the same unit.
This rule immediately covers products from two disjoint edges, products from two edges sharing one unit with the same treatment on that unit, and products from the same unordered pair with consistent treatment requirements on both units.
The last case is unique to edge-level outcomes. Besides squared terms such as $Y^2_{i\to j}(z_1,z_2)$, it also includes reciprocal-edge products such as $Y_{i\to j}(z_1,z_2)Y_{j\to i}(z_2,z_1)$, which have no analogue in standard node-level settings.
Thus, instead of diagonalizing the entire variance expression, we retain the compatible cross-products and bound only the remaining incompatible ones.
This yields a refined bound based on a non-diagonal symmetric matrix that more closely approximates $\tilde{\bm Q}$.
Specifically, define
\begin{align*}
    R_{ij}(z_i,z_j)=\frac{\omega(z_i,z_j)Y_{i\to j}(z_i,z_j)}{\pi(z_i)}+\frac{\omega(z_j,z_i)Y_{j\to i}(z_j,z_i)}{\pi(z_i)}.
\end{align*}
Then
\begin{align*}
    &K_i=\sum_{j=1}^n\left\{R_{ij}(1,1)+R_{ij}(1,0)-R_{ij}(0,1)-R_{ij}(0,0)\right\},\\
    &K_{ij}+K_{ji}=\frac{R_{ij}(1,1)}{r_1}-\frac{R_{ij}(1,0)}{r_0}-\frac{R_{ij}(0,1)}{r_1}+\frac{R_{ij}(0,0)}{r_0}.
\end{align*}
Substituting these expressions into the definition of $\var(\hat \tau_1)$ and $\var(\hat \tau_2)$ yields
\begin{align*}
    \var(\hat \tau_1)&=\frac{r_1r_0}{n^2}\sum_{i=1}^n\left[\sum_{j:j\neq i}\left\{R_{ij}(1,1)+R_{ij}(1,0)-R_{ij}(0,1)-R_{ij}(0,0)\right\}\right]^2,\\
    \var(\hat \tau_2)&=\frac{r_1^2r_0^2}{2n^2}\sum_{(i,j):i\neq j}\left\{\frac{R_{ij}(1,1)}{r_1}-\frac{R_{ij}(1,0)}{r_0}-\frac{R_{ij}(0,1)}{r_1}+\frac{R_{ij}(0,0)}{r_0}\right\}^2.
\end{align*}
We propose two estimable variance targets for $\var(\hat \tau)$. Both are constructed by adding suitable correction terms to preserve conservativeness, while handling the remaining inestimable components differently: bounding these components yields the finite-sample upper bound $\varu 1(\hat \tau)$, whereas omitting them leads to the asymptotically conservative target $\varu 2(\hat \tau)$. Specifically,
\begin{align*}
    \varu 2(\hat \tau)&=\frac {r_1}{n^2}\sum_{i=1}^n\sum_{j\neq j'}\{R_{ij}(1,1)+R_{ij}(1,0)\}\{R_{ij'}(1,1)+R_{ij'}(1,0)\}\\
    &\quad +\frac {r_0}{n^2}\sum_{i=1}^n\sum_{j\neq j'}\{R_{ij}(0,1)+R_{ij}(0,0)\}\{R_{ij'}(0,1)+R_{ij'}(0,0)\}\\
    &\quad +\frac 1{2n^2}\sum_{(i,j):i\neq j}\left[\{R_{ij}(1,1)\}^2+\frac{r_1}{r_0}\{R_{ij}(1,0)\}^2+\frac{r_0}{r_1}\{R_{ij}(0,1)\}^2+\{R_{ij}(0,0)\}^2\right],\\
    \varu 1(\hat \tau)&=\varu 2(\hat \tau)\\
    &\quad +\frac 2{n^2}\sum_{(i,j):i\neq j}\left[r_1^2\{R_{ij}(1,1)\}^2+r_1^2\{R_{ij}(1,0)\}^2+r_0^2\{R_{ij}(0,1)\}^2+r_0^2\{R_{ij}(0,0)\}^2\right].
\end{align*}
Both targets admit unbiased Horvitz--Thompson estimators, with explicit expressions given in \Cref{eq:varhat1,eq:varhat2} of the Supplementary Material. 
\Cref{theorem.variance} summarizes their theoretical properties.

\begin{theorem}
\label{theorem.variance}
Under Assumptions \ref{assumption.bounded}--\ref{assumption.sigma}, we have the following results:
\begin{enumerate}
\item [(i.1)] $\varu 1(\hat \tau) - \var(\hat \tau) \geq 0$ with $\varu 1(\hat \tau) - \var(\hat \tau) = O(n\rho_n^2)$. 
\item [(i.2)] If $n\rho_n \to \infty$ as $n \to \infty$, then $\lim\limits_{n \rightarrow \infty} \{\varu 2(\hat \tau) - \var(\hat \tau)\} \geq 0$ with $|\varu 2(\hat \tau) - \varu 1(\hat \tau)| = O(\rho_n)$.
\item [(ii.1)] $E\{\varhatu 1(\hat \tau)\} = \varu 1(\hat \tau)$ and $E\{ \varhatu 2(\hat \tau) \} = \varu 2(\hat \tau)$.
\item [(ii.2)] $\varhatu 1(\hat \tau) - \varu 1(\hat \tau) = o_p(n\rho_n^2)$ and $\varhatu 2(\hat \tau) - \varu 2(\hat \tau) = o_p(n\rho_n^2)$.
\item [(iii)] The difference between $\varu 2(\hat \tau)$ and $\var(\hat \tau)$ is 
\begin{align*}
&\frac 1{n^2}\sum_{i=1}^n\left[\sum_{j:j\neq i}\left\{r_1R_{ij}(1,1)+r_1R_{ij}(1,0)+r_0R_{ij}(0,1)+r_0R_{ij}(0,0)\right\}\right]^2\\
&\quad -\frac 1{2n^2}\sum_{(i,j):i\neq j}\left\{r_1R_{ij}(1,1)+r_1R_{ij}(1,0)+r_0R_{ij}(0,1)+r_0R_{ij}(0,0)\right\}^2.
\end{align*}
\end{enumerate}
\end{theorem}

\Cref{theorem.variance} establishes theoretical guarantees for variance estimation. 
In particular, $\varhatu 1(\hat{\tau})$ is conservative, generalizing the result of \citet{li2025causal}, who proved this property under the much stronger condition $\max_i m_i \le 2$, and the second estimator $\varhatu 2(\hat \tau)$ is asymptotically conservative under the condition $n\rho_n \to \infty$. 
Moreover, \Cref{theorem.variance}(iii) implies that $\varhatu 2(\hat \tau)/(n\rho_n^2)$ is consistent for $\var(\hat\tau)/(n\rho_n^2)$ if $r_1R_{ij}(1,1) + r_1R_{ij}(1,0) + r_0R_{ij}(0,1) + r_0R_{ij}(0,0) = 0$ for all $i,j$. Rewriting this condition in terms of $\omega$ and the potential outcomes gives
$\sum_{z_i,z_j\in\{0,1\}}\{\omega(z_i,z_j)\Y{i}{j}{z_i,z_j}+\omega(z_j,z_i)\Y{j}{i}{z_j,z_i}\}=0$.

Based on these theoretical results, we can construct (asymptotically) valid confidence intervals using the proposed variance estimators. 
We recommend using the second variance estimator when the average number of network neighbors is large, as it yields shorter confidence intervals while preserving coverage probabilities close to the nominal level.

\section{Covariate adjustment}
\label{sec:covariate}

\subsection{Three-fold sample splitting and cross-fitting}
\label{sec:three-fold}

To improve the statistical efficiency of the Horvitz--Thompson estimator, we employ covariate adjustment, a powerful and widely used approach for variance reduction in randomized experiments \citep{freedman2008regression, lin2013agnostic,bloniarz2016lasso,lei2021regression,liu2020regression,lu2025debiased}.

Let $\bm X_{i\to j}\in \mathbb R^{d\times 1}$ denote a $d$-dimensional covariate vector for the edge $i \to j$. 
These covariates may include node attributes (e.g., age and gender), edge attributes (e.g., past interaction frequency), and network features (e.g., common neighbors and centrality measures). Building on sample splitting and cross-fitting ideas for design-based causal inference with node-level outcomes \citep{su2023decorrelation,lu2025conditional}, we develop a new edge-level generalization that enables the use of flexible machine learning predictions for covariate adjustment while preserving unbiasedness and valid inference.

%In network settings, edge-level potential outcomes are zero for non-edges ($i \to j$ with $A_{ij}=0$). If a prediction model assigns nonzero values to such edges, it introduces additional variance and degrades inference.

In this section, we assume that the adjacency matrix $\bm A$ is known and set $\bm X_{i\to j}=\bm 0$ whenever $A_{ij}=0$ in the covariate adjustment. Our new sample splitting and cross-fitting approach consists of three main steps:

\textbf{Step 1: Data Splitting.} Randomly partition the nodes $\{1,\ldots,n\}$ into three disjoint subsets. 
Specifically, assign each node independently to $\mathcal{S}_1,\mathcal{S}_2,\mathcal{S}_3$ with equal probability $1/3$, so that $\mathcal{S}_1 \cup \mathcal{S}_2 \cup \mathcal{S}_3 = \{1,\ldots,n\}$.

\begin{figure}[!ht]
    \centering
        \includegraphics[width=0.9\linewidth]{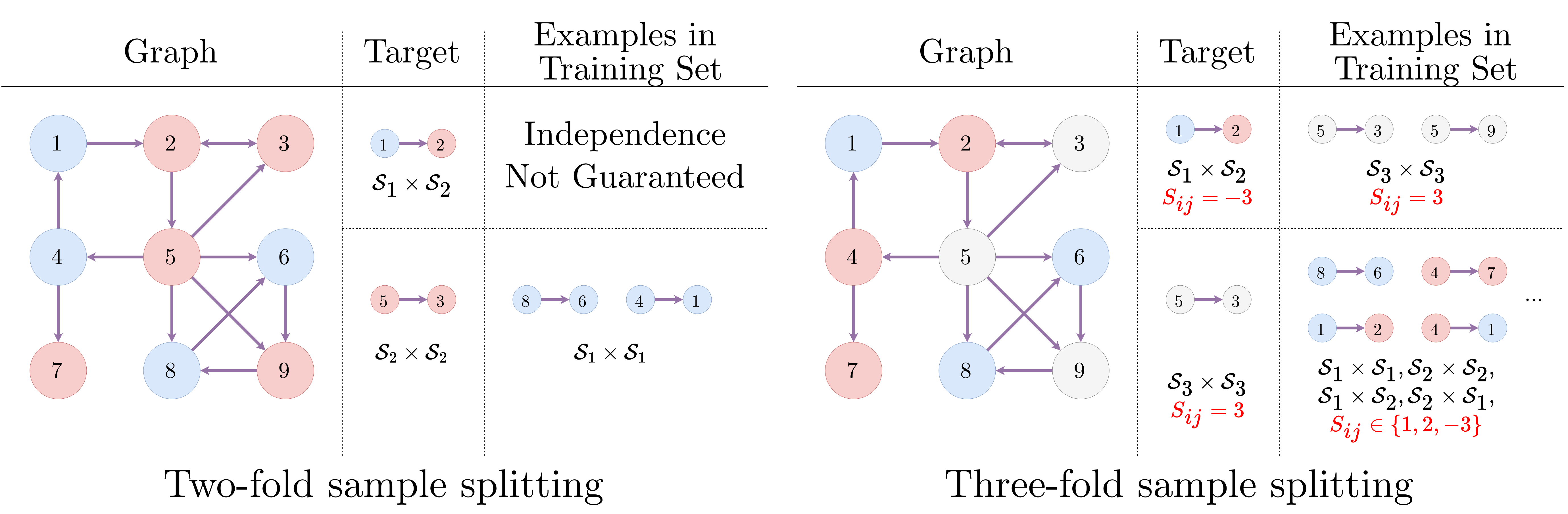}
    \caption{Comparison of two-fold and three-fold sample splitting.\\Nodes with the same color belong to the same fold.}
    \label{fig:fold}
\end{figure}

The standard \textbf{two-fold} splitting approach cannot accommodate the dyadic structure of edge-level outcomes; hence, at least a \textbf{three-fold} splitting strategy is required.
%The \textbf{three-fold} splitting strategy is a necessary refinement of the standard \textbf{two-fold} splitting approach, which cannot accommodate the dyadic structure of edge-level outcomes. 
As illustrated in the first panel of \Cref{fig:fold}, two-fold splitting fails because an edge may connect a sender in one fold to a receiver in the other (e.g., edge $1 \to 2$). 
To ensure exact unbiasedness, the fitted model used for an edge must be trained on data independent of that edge. 
For such cross-fold edges, this would require excluding data from both endpoint folds, leaving no data for training. The three-fold splitting strategy resolves this issue, as shown in the second panel of \Cref{fig:fold}. 
A three-fold partition ensures that each edge falls into one of six categories determined by the subset memberships of its endpoints. 
Specifically, define the indicator
\begin{align*}
    S_{ij} =
    \begin{cases}
    k & \text{if } i, j \in \mathcal{S}_k, \quad k = 1,2,3, \\
    -1 & \text{if } (i, j) \in \mathcal{S}_2 \times \mathcal{S}_3 \cup \mathcal{S}_3 \times \mathcal{S}_2, \\
    -2 & \text{if } (i, j) \in \mathcal{S}_1 \times \mathcal{S}_3 \cup \mathcal{S}_3 \times \mathcal{S}_1, \\
    -3 & \text{if } (i, j) \in \mathcal{S}_1 \times \mathcal{S}_2 \cup \mathcal{S}_2 \times \mathcal{S}_1.
    \end{cases}
\end{align*}
This labeling partitions edges into three \textbf{within-fold} categories ($S_{ij} \in \{1,2,3\}$, e.g., $5\to 3$) and three \textbf{between-fold} categories ($S_{ij} \in \{-1,-2,-3\}$, e.g., $1\to 2$), which facilitates model training and evaluation across all edge types.

\textbf{Step 2: Model Fitting.} 
The fitted models and their corresponding training sets are defined as follows:
\begin{align*}
\hat{f}_{\z,1} &\text{ is trained on } \{(i,j) : S_{ij} \in \{-1, 2, 3\}\}, \
\hat{f}_{\z,2} \text{ is trained on } \{(i,j) : S_{ij} \in \{1, -2, 3\}\}, \\
\hat{f}_{\z,3} &\text{ is trained on } \{(i,j) : S_{ij} \in \{1, 2, -3\}\}, \
\hat{f}_{\z,-k} \text{ is trained on } \{(i,j) : S_{ij} = k\}, \ k = 1, 2, 3.
\end{align*}
It is guaranteed that the model $\hat{f}_{\z, S_{ij}}(\cdot)$ used to predict the outcome on edge $(i, j)$ is trained on data from edges whose endpoints do \textbf{not} share a fold with either node $i$ or node $j$.

\textbf{Step 3: Estimator Construction.} We construct the covariate-adjusted estimator for the average potential outcome under treatment configuration $\z$ as follows:
\begin{align*}
\hat{\theta}^{\adj}(\z) = \frac{1}{n} \sum_{(i,j):i\neq j} A_{ij}  \left[\frac{I_{ij}(\z)}{\pi(\z)} \left\{\Yo{i}{j} - \hat{f}_{\z,S_{ij}}(\bm X_{i\to j})\right\} + \hat{f}_{\z,S_{ij}}(\bm X_{i\to j})\right].
\end{align*}
The estimator for the average causal effect $\tau$ is then constructed as the weighted sum $\hat{\tau}^{\adj} = \sum_{\z} \omega(\z) \hat{\theta}^{\adj}(\z)$. This can be recognized as an AIPW-type estimator adapted to edge-level outcomes.

\subsection{Properties of the covariate-adjusted estimator}

\Cref{theorem.adjunbiased} below shows that the covariate-adjusted estimator $\hat \tau^{\adj}$ is unbiased for $\tau$ conditional on the random split, which further implies marginal unbiasedness.

\begin{theorem}
    \label{theorem.adjunbiased}
    $E(\hat \tau^{\adj}\mid \mathcal S_1,\mathcal S_2,\mathcal S_3)=\tau$, and consequently $E(\hat \tau^{\adj})=\tau$.
\end{theorem}

To establish asymptotic normality of the covariate-adjusted estimator, we introduce the oracle function $f_{\z}^{\ora}$, an \emph{infeasible} population-level target defined for each treatment configuration $\z$ using the full set of potential outcomes. We provide two examples in Section~\ref{sec:linear}.
The corresponding oracle estimator is defined as
\begin{align*}
    \hat \theta^{\ora}(\z)=\frac{1}{n}\sum_{(i,j):i\neq j} A_{ij}\left[\frac{I_{ij}(\z)}{\pi(\z)}\left\{\Y{i}{j}{\z}-f_{\z}^{\ora}(\bm X_{i\to j})\right\}+f_{\z}^{\ora}(\bm X_{i\to j})\right],
\end{align*}
and the oracle estimator for the average causal effect is $\hat \tau^{\ora}=\sum_{\z}\omega(\z)\hat \theta^{\ora}(\z)$.
The fitted functions obtained from the three-step procedure introduced above serve as approximations to the oracle function.
If they converge to the oracle functions at a sufficiently fast rate, the resulting estimator $\hat \tau^{\adj}$ is asymptotically equivalent to $\hat \tau^{\ora}$ and inherits its asymptotic normality, as demonstrated in \Cref{theorem.equivalence} below.

Let $\e{i}{j}{\z} = \Y{i}{j}{\z} - f_{\z}^{\ora}(\bm X_{i\to j})$ denote the oracle residuals. 
Define $\sigora^2$ as the analogue of $\sig^2$ with $\Y{i}{j}{\z}$ replaced by $\e{i}{j}{\z}$.
The following assumption parallels Assumptions~\ref{assumption.bounded} and \ref{assumption.sigma}, formulated for the oracle residuals $\e{i}{j}{\z}$.

\begin{assumption}
    \label{assumption.covariate}
        (i) The residuals \(\e{i}{j}{\z}\) are uniformly bounded, meaning that there exists a constant $C$ such that $|\e{i}{j}{\z}| \leq C$ for all $1\leq i\neq j\leq n$ and $\z\in\{0,1\}^2$. (ii) $\liminf_{n \to \infty} \sigora^2/(n \rho_n^2) > 0$.
\end{assumption}

\Cref{assumption.stability} below imposes a stability condition requiring the fitted functions $\hat f_{\z,k}$ and $\hat f_{\z,-k}$ to converge sufficiently fast to the oracle function $f_{\z}^{\ora}$. 
This condition is essential for establishing the asymptotic normality of the covariate-adjusted estimator $\hat \tau^{\adj}$. 

\begin{assumption}[Stability]
\label{assumption.stability}
For $\z\in\{0,1\}^2$ and $\hat{\varepsilon}_{\z,ij} = \hat{f}_{\z,S_{ij}}(\bm X_{i\to j}) - f_{\z}^{\ora}(\bm X_{i\to j})$, we have
\begin{align*}
    \frac{\sum_{(i,j):i\neq j}A_{ij}(m_i+m_j) \hat \varepsilon_{\z,ij}^2}{n^3\rho_n^2} = o_p(1).
\end{align*}
\end{assumption}

Since $(n^3\rho_n^2)^{-1}\sum_{(i,j):i\neq j} A_{ij}(m_i+m_j) \{\Y{i}{j}{\z}\}^2 = O(1)$, \Cref{assumption.stability} ensures that the estimation errors are asymptotically negligible relative to the potential outcomes $\Y{i}{j}{\z}$. 
This condition is similar to stability assumptions in the literature \citep{guo2023generalized, cohen2024no}, but is tailored to the edge-level potential outcome framework.

\begin{theorem}
    \label{theorem.equivalence}
    Under Assumptions \ref{assumption.bounded}--\ref{assumption.stability}, $\{\var(\hat \tau^{\ora})\}^{-1/2}(\hat \tau^{\adj}-\hat \tau^{\ora})=o_p(1)$, and $(\hat \tau^{\adj}-\tau)/\sqrt{\var(\hat \tau^{\ora})}\xrightarrow{d}\mathcal N(0,1)$ as $n\to \infty$.
\end{theorem}

\Cref{theorem.equivalence} shows that $\hat{\tau}^{\adj}$ is asymptotically equivalent to $\hat{\tau}^{\ora}$ and is therefore asymptotically normal.  
To enable valid inference, we introduce variance estimators for $\hat \tau^{\adj}$. 
Let $\ehat{i}{j}{\z} = \Y{i}{j}{\z} - \hat f_{\z,S_{ij}}(\bm X_{i\to j})$ denote the residuals based on the fitted functions.
We define two variance estimators $\varhatu 1(\hat \tau^{\adj})$ and $\varhatu 2(\hat \tau^{\adj})$ by replacing $\Y{i}{j}{\z}$ with $\ehat{i}{j}{\z}$ in $\varhatu 1(\hat \tau)$ and $\varhatu 2(\hat \tau)$. 

\begin{theorem}
    \label{theorem.covariateVariance}
    Under Assumptions \ref{assumption.bounded}--\ref{assumption.stability}, we have

    (i) $\varhatu 1(\hat \tau^{\adj})-\varu 1(\hat \tau^{\ora})= o_p (n\rho_n^2)$ and $\varu 1(\hat \tau^{\ora})-\var(\hat \tau^{\ora})\geq 0$;

    (ii) $\varhatu 2(\hat \tau^{\adj})-\varu 2(\hat \tau^{\ora})=o_p(n\rho_n^2)$ and $\lim\limits_{n \rightarrow \infty}\{\varu 2(\hat \tau^{\ora})-\var(\hat \tau^{\ora})\}\geq 0$, provided that $n\rho_n\to \infty$ as $n\to \infty$.
\end{theorem}

\Cref{theorem.covariateVariance} shows that $\varhatu 1(\hat \tau^{\adj})$ is conservative and that $\varhatu 2(\hat \tau^{\adj})$ is asymptotically conservative when $n\rho_n \to \infty$, paralleling \Cref{theorem.variance}. 
These variance estimators enable the construction of (asymptotically) valid confidence intervals based on the covariate-adjusted estimator $\hat \tau^{\adj}$. 
Analogous to the unadjusted case, we recommend using the second variance estimator when the average degree of the network is large.

\subsection{Linear adjustment and beyond}
\label{sec:linear}

In our proposed covariate-adjustment framework, the choice of model is flexible, ranging from linear models to more complex machine learning methods. 
Below, we first present linear models as a primary approach, followed by a discussion of more general prediction methods based on loss functions. 
In the following, we slightly abuse notation by writing $f(\bm X)$ for the $n^2\times 1$ vector of fitted values, where $\bm X$ is the $n^2\times d$ covariate matrix with rows $\bm X_{(i-1)n+j,\cdot} = \bm X_{i\to j}$.  
The $\{(i-1)n+j\}$-th entry of $f(\bm X)$ corresponds to edge $i\to j$ and is defined as $f(\bm X_{i\to j})$.  
The same convention is used for both oracle and fitted functions.

\textbf{Linear models}. Linear regression adjustment is a basic and powerful tool for variance reduction in the node-level outcome framework \citep{freedman2008regression,lin2013agnostic}. 
In this setting, the oracle function takes the linear form $f^{\ora}_{\z}(\bm X_{i\to j})=\bm X_{i\to j}^{\top}\bm \beta^{\ora}_{\z}$, where $\bm{\beta}^{\ora}_{\z} \in \mathbb{R}^{d\times 1}$ is the oracle coefficient vector. 
The covariates may include a vector of ones, allowing the model to incorporate an intercept. 
To ensure that covariate adjustment does not increase variance, i.e., satisfies the ``do-no-harm'' property \citep{guo2023generalized}, we define the oracle coefficient as the solution to a generalized least squares (GLS) problem.
Specifically, $\bm \beta^{\ora}=[(\bm \beta^{\ora}_{(1,1)})^{\top},(\bm \beta^{\ora}_{(1,0)})^{\top},(\bm \beta^{\ora}_{(0,1)})^{\top},(\bm \beta^{\ora}_{(0,0)})^{\top}]^{\top}$ is defined as $\bm \beta^{\ora}=\argmin_{\bm \beta}(\bm c^{\mathrm{LR}})^{\top} \tilde{\bm Q} (\bm c^{\mathrm{LR}})$,
where 
\begin{align}
\label{eqn:CLR}
    \bm c^{\mathrm{LR}}_{4(i-1)n+4(j-1)+1:4}=\begin{bmatrix}
        \omega(1,1)\{\Y{i}{j}{1,1}-\bm X_{i\to j}^{\top}\bm \beta_{(1,1)}\}\\
        \omega(1,0)\{\Y{i}{j}{1,0}-\bm X_{i\to j}^{\top}\bm \beta_{(1,0)}\}\\
        \omega(0,1)\{\Y{i}{j}{0,1}-\bm X_{i\to j}^{\top}\bm \beta_{(0,1)}\}\\
        \omega(0,0)\{\Y{i}{j}{0,0}-\bm X_{i\to j}^{\top}\bm \beta_{(0,0)}\}
    \end{bmatrix}.
\end{align}
The explicit expression for $\bm \beta^{\ora}$ is given in \Cref{eq:beta} of the Supplementary Material.
This GLS method ensures that the covariate-adjusted estimator has variance no larger than that of the unadjusted estimator $\hat \tau$, thereby guaranteeing efficiency gains, as shown in \Cref{theorem.donoharm}.

\begin{theorem}
    \label{theorem.donoharm}
    Under Assumptions \ref{assumption.bounded}--\ref{assumption.sigma}, for linear models, we have $\var(\hat \tau^{\ora}) \leq \var(\hat \tau)$.
\end{theorem}

Since the oracle functions $f^{\ora}_{\z}$ are unobservable, they are estimated from the observed data using the three-step procedure introduced in \Cref{sec:three-fold}. The fitted function $\hat f_{\z,k}$ is specified as $\hat f_{\z,k}(\bm X_{i\to j})=\bm X_{i\to j}^{\top}\hat{\bm \beta}_{\z,k}$, where $\hat{\bm \beta}_{\z,k}$ is the inverse probability weighted estimator of $\bm \beta^{\ora}_{\z}$. 
Specifically, $Y_{i\to j}(\z)$ in the definition of $\bm \beta^{\ora}_{\z}$ is replaced by its inverse probability weighted counterpart, with weights given by
\begin{align*}
    &\frac {I(S_{ij}\in\{-1,2,3\})}{\pr(S_{ij}\in\{-1,2,3\})}\frac{I_{ij}(\z)}{\pi(\z)} \text{ for } \hat{\bm \beta}_{\z,1},\ 
    \frac {I(S_{ij}\in\{1,-2,3\})}{\pr(S_{ij}\in\{1,-2,3\})}\frac{I_{ij}(\z)}{\pi(\z)} \text{ for } \hat{\bm \beta}_{\z,2},\\
    &\frac {I(S_{ij}\in\{1,2,-3\})}{\pr(S_{ij}\in\{1,2,-3\})}\frac{I_{ij}(\z)}{\pi(\z)} \text{ for } \hat{\bm \beta}_{\z,3},\ 
    \frac {I(S_{ij}=k)}{\pr(S_{ij}=k)}\frac{I_{ij}(\z)}{\pi(\z)} \text{ for } \hat{\bm \beta}_{\z,-k},\ k=1,2,3.
\end{align*}
Since $\bm \beta_{\z}^{\ora}$ is linear in the potential outcomes, the fitted coefficients are unbiased for $\bm \beta_{\z}^{\ora}$.
We next show that \Cref{assumption.stability} holds for the fitted linear models $\hat f_{\z,k}$ under suitable conditions, as formalized in \Cref{theorem.linear}, even when the number of covariates diverges.

\begin{theorem}
\label{theorem.linear}
Suppose that Assumptions~\ref{assumption.bounded}--\ref{assumption.sigma} hold, and that 
\begin{align*}
& \max_i m_i=O(1),\quad
d = o(n), 
\quad \|(\bm M^{\top}\tilde{\bm Q}\bm M)^{+}\|_{2}=O(m^{-1}),\\
&  \lambda_{\max}\bigg( (n^{3}\rho_{n}^{2})^{-1} \Big\{
\sum_{(i,j):i\neq j}(m_i+m_j)\bm X_{i\to j}\bm X_{i\to j}^{\top}\Big\} \bigg) =O(1),
\end{align*}
where $\bm M = \bm I_4 \otimes \bm X$ with $\bm I_4$ denoting the $4\times4$ identity matrix and $\otimes$ denoting the Kronecker product, and $(\cdot)^+$ denotes the Moore--Penrose pseudo-inverse.
Then, Assumption~\ref{assumption.stability} holds.
\end{theorem}

\Cref{theorem.linear} establishes explicit technical conditions that guarantee \Cref{assumption.stability} holds. These conditions are satisfied with high probability when $m_i$ is bounded, $\bm X_{i\to j}$ is drawn from a zero-mean distribution and the covariance between $\bm X_{i\to j}$ and $\bm X_{i'\to j'}$ is determined solely by the number of shared endpoints ($0$, $1$, or $2$). Specifically, edges with no shared endpoints have independent covariates, while the covariance matrix for edges sharing endpoints has eigenvalues bounded away from zero and infinity.
The dimensionality requirement, $d=o(n)$, is considerably weaker than the conditions in the classical literature \citep{lin2013agnostic,lei2021regression} and aligns with recent advances in linear adjustment under diverging-dimensional settings \citep{su2023decorrelation,lu2025debiased}.

\textbf{Nonparametric or machine learning models}. 
Beyond linear models, our framework accommodates general prediction methods, including nonparametric and machine learning approaches. For a given function class $\mathcal F_{\z}$, define the oracle function $f_{\z}^{\ora}$ as the minimizer of the loss
$f_{\z}^{\ora}= \argmin_{f\in\mathcal F_{\z}}[m^{-1}\sum_{(i,j):i\neq j}l_{\z}\{\bm X_{i\to j}, \Y{i}{j}{\z}, f\}]$.
For example, in a bounded nonparametric regression setting, one may take
$l_{\z}\{\bm X_{i\to j}, \Y{i}{j}{\z}, f\}=A_{ij}\{f(\bm X_{i\to j})-\Y{i}{j}{\z}\}^2$.
The fitted functions $\hat f_{\z,k}$ and $\hat f_{\z,-k}$ are then obtained by minimizing the corresponding empirical loss computed with the observed outcomes. Under suitable regularity conditions, we establish non-asymptotic error bounds for the fitted functions and verify that they satisfy \Cref{assumption.stability}. Details are provided in \Cref{sec.sm:nonasymptotic} of the Supplementary Material.

\subsection{Calibration and refinement}
\label{sec:calibration}

While the covariate adjustment method allows for flexible predictive models, such models do not automatically guarantee variance reduction, unlike the linear adjustment method in \Cref{sec:linear}.
To address this issue, we introduce a \textbf{two-step} calibration and refinement procedure, inspired by \citet{cohen2024no}, that guarantees the covariate-adjusted estimator has asymptotic variance no larger than that of the unadjusted estimator.
Specifically, we replace Step~2 in \Cref{sec:three-fold} with a two-step procedure: first fitting a prediction model, and then applying the linear adjustment proposed in \Cref{sec:linear} using the fitted values obtained from the prediction model as covariates.
Technical details are provided in \Cref{sec.sm:calibration} of the Supplementary Material.

\section{Simulation}
\label{sec:simulation}

This section evaluates the performance of the edge-level causal inference methods through simulations.  
We consider networks of varying population sizes ($n$), sparsity levels ($m/n$), treatment proportions ($r_1$), and treatment-configuration weights ($\bm{\omega}$).  
Potential outcomes are generated under multiple data-generating processes (DGPs) to assess robustness across diverse scenarios.  
The network is generated using an Erd\H{o}s--R\'enyi graph. Performance is measured in terms of bias, variance, mean squared error (MSE), and empirical coverage probability (CP) of $95\%$ confidence intervals. Specifically, we consider the following setup:
    
\textbf{Scenario 1:} Varying network size $n\in\{200,500,1000\}$, fixed $m/n=10$, fixed treated proportion $r_1=0.5$, and fixed $\bm \omega=(1,0,0,-1)$, which corresponds to the total effect.

\textbf{Scenario 2:} Fixed network size $n=500$, varying $m/n\in\{10,20,30\}$, fixed treated proportion $r_1=0.5$, and fixed $\bm \omega=(1,0,0,-1)$.

\textbf{Scenario 3:} Fixed network size $n=500$, fixed $m/n=10$, varying treated proportion $r_1\in\{0.3,0.5,0.7\}$, and fixed $\bm \omega=(1,0,0,-1)$.

\textbf{Scenario 4:} Fixed network size $n=500$, fixed $m/n=10$, fixed treated proportion $r_1=0.5$, and varying $\bm \omega$: $\bm{\omega} \in \{(0,1,0,-1), (0,0,1,-1), (1,0,0,-1)\}$, corresponding to the sender effect, receiver effect, and total effect, respectively.  

\textbf{Covariate and DGP:} Each edge $i \to j$ is associated with a covariate vector $\bm X_{i\to j} = (X_{i\to j,1}, X_{i\to j,2})$, where each component is independently drawn from $\text{Uniform}([-2, 2])$. For each edge $i \to j$, four independent parameters $\varphi_{ij,1}, \varphi_{ij,2}, \varphi_{ij,3}, \varphi_{ij,4} \overset{\text{i.i.d.}}{\sim} \mathcal{N}(0,1)$ are generated. These parameters are then used to define $\mu_{ij}(z_i, z_j) = \varphi_{ij,1} + (2z_i - 1)\varphi_{ij,2} + (2z_j - 1)\varphi_{ij,3} + (2z_i - 1)(2z_j - 1)\varphi_{ij,4}$. The full potential outcome for each edge is constructed under the following three DGPs:

\textbf{DGP 1:} $\Y{i}{j}{\z} = X_{i\to j,1}^2 + X_{i\to j,2}^2 + \mu_{ij}(\z)$;

\textbf{DGP 2:} $\Y{i}{j}{\z} = 1 + X_{i\to j,1} + X_{i\to j,2} + \mu_{ij}(\z)$;

\textbf{DGP 3:} $\Y{i}{j}{\z} = X_{i\to j,1}^2 + X_{i\to j,2}^2 + X_{i\to j,1} \cdot X_{i\to j,2} + \mu_{ij}(\z)$.

We present the results for DGP 1 under Scenarios 1 and 2 in the main text, and defer the remaining results to \Cref{sec.sm:simulation} of the Supplementary Material. Figures~\ref{fig:1} and~\ref{fig:2} display the simulation results for Scenarios 1 and 2, respectively. Each figure includes two subplots: the first displays the violin plot of the estimator, while the second shows the variance along with the results for two variance estimators. We compare four methods: the baseline estimator without covariate adjustment (labeled as ``unadj''), the linear regression-based covariate-adjusted estimator (labeled as ``LR''), the B-spline-based covariate-adjusted estimator (labeled as ``BS''), and the random forest-based covariate-adjusted estimator (labeled as ``RF''). For BS and RF, we apply the calibration procedure with $(\bm f^{\ora})_{(i-1)n+j,\cdot}=[ 1 \quad f_{\z}^{\ora}(\bm X_{i\to j}) ]$, which reduces the dimensionality of the calibration design matrix and improves computational efficiency.

In the first subfigure of each figure, the differences between the estimated and true average causal effects are centered around zero, confirming that the estimators are unbiased. The second subfigure shows that the variances of the covariate-adjusted estimators are consistently lower than that of the unadjusted estimator, confirming the efficiency gains associated with covariate adjustment. When the DGP is nonlinear, both the BS and RF methods exhibit lower variance than LR in most scenarios, highlighting the effectiveness of nonlinear methods in capturing complex relationships within the data. However, when the sample size is \( n = 200 \), the BS method may perform worse than LR, potentially due to the insufficient sample size for BS to converge. Under DGP 1, excluding \( n = 200 \), the BS and RF methods with calibration reduce the variance by an average of 9.4\% and 10.2\%, respectively, compared with the LR estimator. Conversely, when the DGP primarily consists of linear components, BS and RF may not outperform LR (e.g., DGP 2), as discussed in \Cref{sec.sm:simulation} of the Supplementary Material.

Detailed simulation results for DGP 1 under Scenarios 1 and 2 are reported in \Cref{table:simulation}. 
The table summarizes bias, variance, MSE, two variance estimators, and the empirical CPs of $95\%$ confidence intervals constructed using each variance estimator. 
The results show that covariate-adjusted estimators consistently achieve substantial variance reduction relative to the unadjusted estimator, while maintaining CPs close to the nominal $95\%$ level. 
Moreover, the second variance estimator yields CPs that are systematically closer to the nominal level than those based on the first, in line with the theoretical guarantees.

\begin{figure}[htbp]
    \centering
    \begin{subfigure}{0.9\linewidth}
        \centering
        \includegraphics[width=\linewidth]{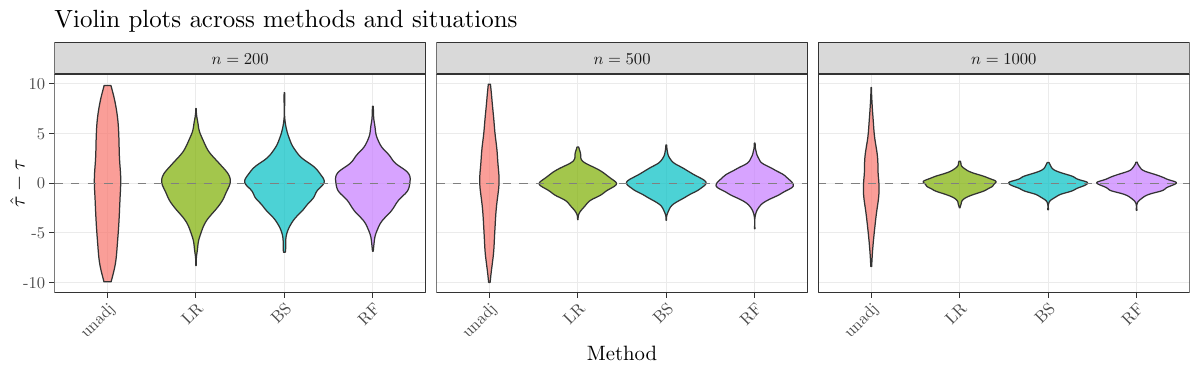}
    \end{subfigure} \\
    \begin{subfigure}{0.9\linewidth}
        \centering
        \includegraphics[width=\linewidth]{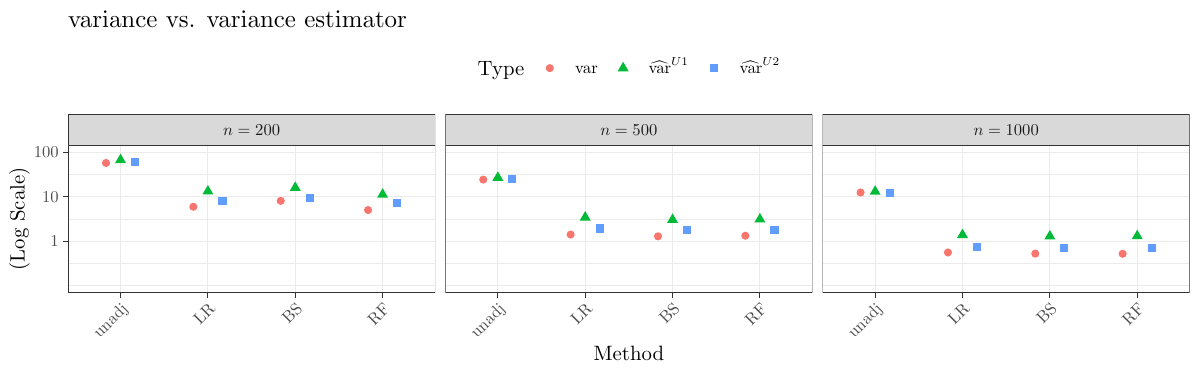}
    \end{subfigure} 
    \caption{Bias and variance for DGP 1 in Scenario 1.}
    \label{fig:1}
\end{figure}

\begin{figure}[htbp]
    \centering
    \begin{subfigure}{0.9\linewidth}
        \centering
        \includegraphics[width=\linewidth]{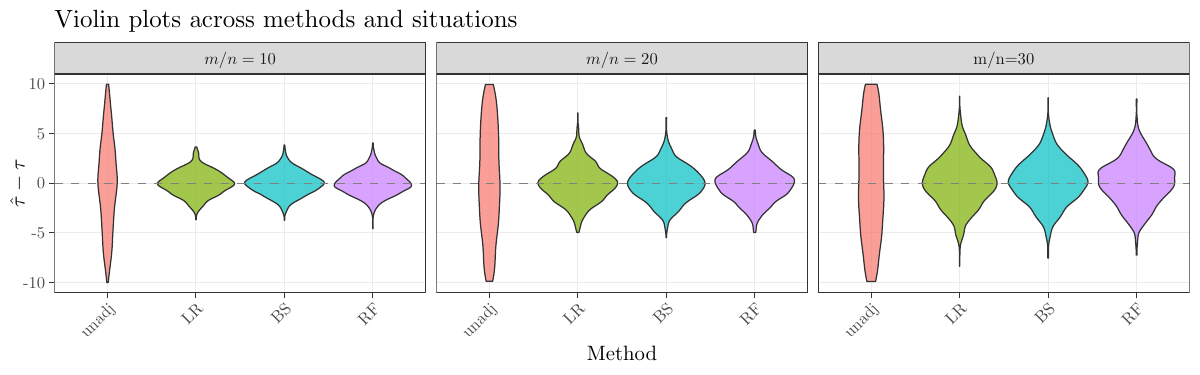}
    \end{subfigure} \\
    \begin{subfigure}{0.9\linewidth}
        \centering
        \includegraphics[width=\linewidth]{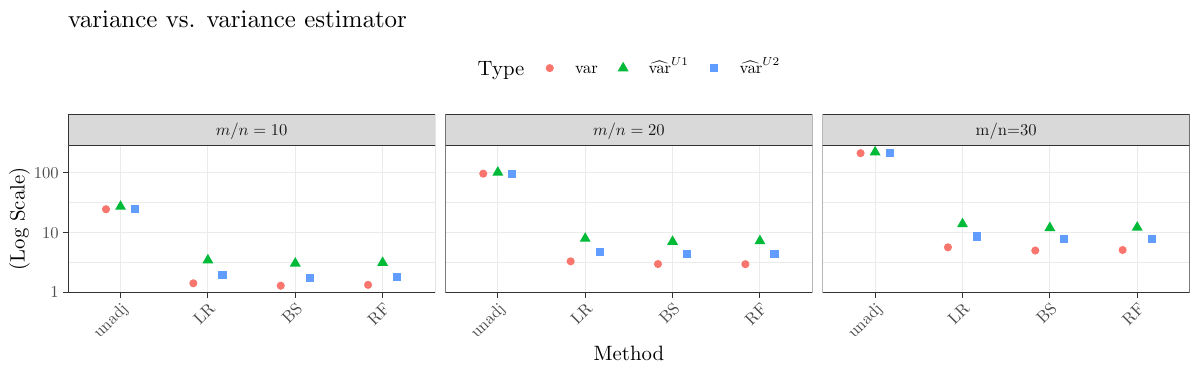}
    \end{subfigure}
    \caption{Bias and variance for DGP 1 in Scenario 2.}
    \label{fig:2}
\end{figure}

\begin{table}[htbp]
\caption{\label{table:simulation}Simulation results for DGP 1 in Scenarios 1 and 2.}
\centering
\begin{tabular}[t]{cccccccccc}
\toprule
$n$ & $m/n$ & Method & Bias & var & MSE & $\varhatu 1$ & CP-1 & $\varhatu 2$ & CP-2\\
\midrule
200 & 10 & unadj & 0.046 & 57.264 & 57.319 & 66.792 & 0.973 & 61.152 & 0.964\\
200 & 10 & LR & -0.108 & 5.948 & 5.942 & 13.206 & 0.998 & 8.140 & 0.973\\
200 & 10 & BS & 0.005 & 8.066 & 8.074 & 15.933 & 0.999 & 9.265 & 0.981\\
200 & 10 & RF & -0.022 & 5.007 & 5.012 & 11.195 & 0.995 & 7.136 & 0.972\\
\addlinespace
500 & 10 & unadj & 0.119 & 24.217 & 24.227 & 26.864 & 0.974 & 24.612 & 0.965\\
500 & 10 & LR & 0.000 & 1.408 & 1.409 & 3.425 & 0.997 & 1.929 & 0.973\\
500 & 10 & BS & -0.004 & 1.284 & 1.285 & 3.032 & 0.995 & 1.745 & 0.970\\
500 & 10 & RF & 0.001 & 1.325 & 1.326 & 3.110 & 0.994 & 1.786 & 0.969\\
\addlinespace
1000 & 10 & unadj & 0.150 & 12.422 & 12.412 & 13.048 & 0.957 & 11.955 & 0.946\\
1000 & 10 & LR & -0.050 & 0.560 & 0.558 & 1.385 & 0.999 & 0.751 & 0.973\\
1000 & 10 & BS & -0.034 & 0.526 & 0.526 & 1.292 & 0.998 & 0.704 & 0.976\\
1000 & 10 & RF & -0.038 & 0.522 & 0.521 & 1.304 & 0.998 & 0.711 & 0.979\\
\addlinespace
500 & 20 & unadj & 0.182 & 95.028 & 95.090 & 99.749 & 0.962 & 95.186 & 0.959\\
500 & 20 & LR & -0.011 & 3.274 & 3.277 & 7.856 & 0.999 & 4.716 & 0.976\\
500 & 20 & BS & -0.023 & 2.951 & 2.954 & 6.978 & 0.998 & 4.280 & 0.978\\
500 & 20 & RF & -0.026 & 2.934 & 2.936 & 7.151 & 1.000 & 4.366 & 0.984\\
\addlinespace
500 & 30 & unadj & 0.497 & 208.884 & 208.846 & 216.890 & 0.956 & 209.989 & 0.954\\
500 & 30 & LR & 0.150 & 5.635 & 5.618 & 13.724 & 0.998 & 8.526 & 0.983\\
500 & 30 & BS & 0.164 & 4.986 & 4.964 & 11.853 & 0.996 & 7.643 & 0.981\\
500 & 30 & RF & 0.138 & 5.076 & 5.062 & 12.040 & 0.996 & 7.735 & 0.983\\
\bottomrule
\end{tabular}
\end{table}

\begin{figure}[!ht]
    \centering
    \begin{subfigure}[b]{0.48\linewidth}
        \centering
        \includegraphics[width=\linewidth]{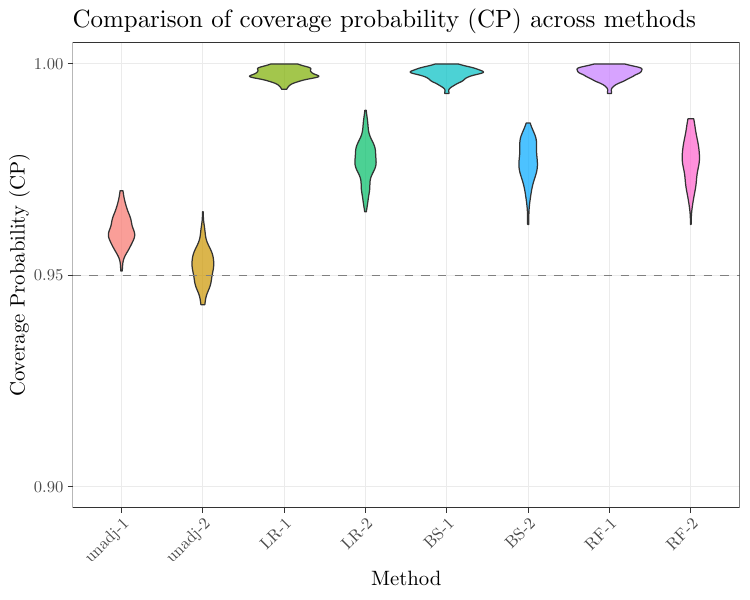}
    \end{subfigure}
    \hfill
    \begin{subfigure}[b]{0.48\linewidth}
        \centering
        \includegraphics[width=\linewidth]{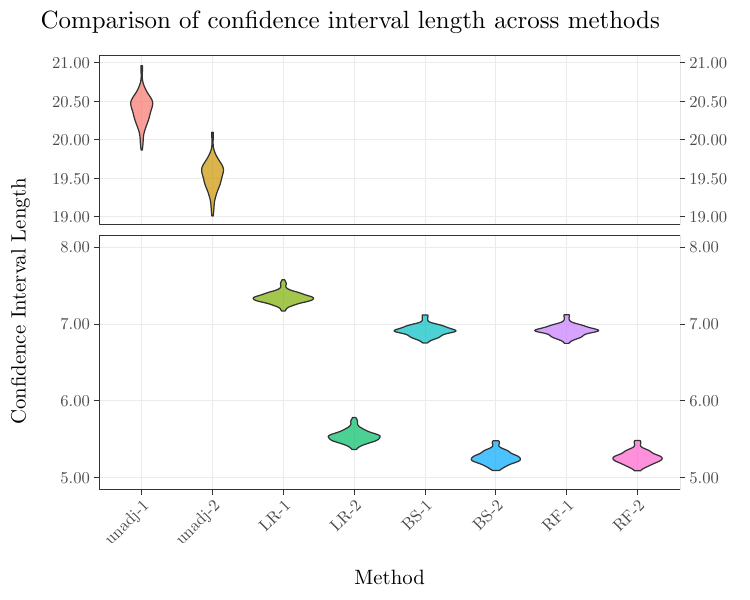}
    \end{subfigure}
    \caption{Empirical coverage probability (CP) and mean confidence interval length\\ based on 100 datasets generated from DGP 1.}
    \label{fig:CP}
\end{figure}

The previous simulation results are design-based, where a single dataset is fixed and treatment assignments are repeatedly randomized. 
To assess robustness to data realization, we generate 100 independent datasets with $n=500$, $m/n=10$, and $r_1=0.5$ under DGP 1. 
Figure~\ref{fig:CP} reports the empirical CPs of confidence intervals based on the unadjusted, LR, BS, and RF estimators for the total effect, where the suffixes ``-1'' and ``-2'' denote the two variance estimators introduced in \Cref{sec:inference}. 
Across all methods and both variance estimators, the empirical CPs are close to or above the nominal $95\%$ level, confirming the validity of the proposed inference procedure. Moreover, the adjusted estimators yield substantially shorter confidence intervals than the unadjusted estimator, demonstrating clear efficiency gains from covariate adjustment. Among the adjusted methods, inference based on $\varhatu{2}$ consistently produces shorter confidence intervals than inference based on $\varhatu{1}$, reflecting the reduced conservativeness of the second variance estimator.

We further assess the necessity of the calibration step in the non-linear covariate adjustment procedure by comparing the estimator's performance with and without calibration.  
The results show that BS and RF without calibration can exhibit increased variance.  
Additional details are provided in \Cref{sec.sm:simulation} of the Supplementary Material.

\section{Real data analysis}
\label{sec:realdata}

We illustrate the proposed methods using an A/B test conducted on WeChat Channels, a content-sharing service within WeChat that allows users to create, share, and discover videos and live streams beyond their friendship network.
In this experiment, treated users received an updated fine-ranking formula that changed the relative weights of several ranking signals, including social-tie, following, thumb-up, and cold-start posting features; control users continued to receive recommendations generated by the previous fine-ranking formula.
For commercial confidentiality, we cannot disclose the precise form of the ranking formula or the exact definitions of the ranking signals.
All data were collected with user authorization and were de-identified before analysis.

The experiment enrolled a simple random sample comprising $n=\{2\%\text{ of all users}\}$ and used Bernoulli randomization with treatment probability $r=0.5$.
We define the adjacency matrix $\bm A$ using the one-to-one chat friendship graph observed during the $30$ days before the experiment.
Specifically, $A_{ij}=1$ if users $i$ and $j$ had at least one private chat interaction in this pre-experiment window, and $A_{ij}=0$ otherwise.
The resulting graph contains billions of directed edges.
For each edge $i\to j$, the observed outcome $\Yo{i}{j}$ measures sender-to-receiver interaction intensity during the experimental period, such as sharing duration, sharing impressions, or successful sharing events from $i$ to $j$.
For the outcome reported below, we construct the edge-level covariate vector $\bm X_{i\to j}$ from three corresponding pre-experiment outcomes, denoted by $Y_{q,i\to j}^{\mathrm{pre}}$ for $q=1,2,3$, measured on the same directed edge during the $14$ days before the experiment.

We focus on the total effect $\tau_{\mathrm{tot}}=\theta(1,1)-\theta(0,0)$,
corresponding to the weight vector $\bm \omega=(1,0,0,-1)$ in \Cref{sec:setting}, and test the null hypothesis $\mathbb H_0:\tau_{\mathrm{tot}}=0$.
Three estimators are implemented: the unadjusted Horvitz--Thompson estimator $\hat\tau^{\mathrm{unadj}}=\hat\tau$ (labeled ``unadj''), the linear-regression-adjusted estimator $\hat\tau^{\mathrm{LR}}$ introduced in \Cref{sec:linear} (labeled ``LR''), and the random-forest-adjusted estimator $\hat\tau^{\mathrm{RF}}$ with the calibration step in \Cref{sec:calibration} (labeled ``RF'').
For each method $*\in\{\mathrm{unadj},\mathrm{LR},\mathrm{RF}\}$ and each variance estimator $\ell=1,2$, inference is based on $\varhatu{\ell}(\hat\tau^{*})$, and the two-sided $p$-value is computed as $2[1-\Phi\{|\hat\tau^{*}|/\{\varhatu{\ell}(\hat\tau^{*})\}^{1/2}\}]$, where $\Phi(\cdot)$ denotes the standard normal CDF. 

\begin{table}[!ht]
\centering
\caption{Relative difference (RD), standard errors (se), and the corresponding two-sided $p$-values for testing $\mathbb H_0:\tau_{\mathrm{tot}}=0$.}
\label{tab:realdata}
\begin{tabular}{lccccc}
\toprule
Method & RD & se-1 & $p$-value-1 & se-2 & $p$-value-2\\
\midrule
unadj
& 1.13\% 
& 0.76\%
& 0.137 
& 0.72\% 
& 0.117 \\

LR
& 1.17\% 
& 0.71\% 
& 0.099 
& 0.69\% 
& 0.090 \\

RF
& 1.24\%
& 0.66\%
& 0.059
& 0.63\% 
& 0.049 \\
\bottomrule
\end{tabular}
\end{table} 

For commercial confidentiality and user privacy protection, the absolute levels of the outcome cannot be disclosed. 
Therefore, \Cref{tab:realdata} reports point estimates and standard errors (se) on the relative-difference scale, obtained by dividing both quantities by $\hat\theta(0,0)$, together with the corresponding $p$-values for the three estimators. 
The suffixes ``-1'' and ``-2'' indicate inference based on $\varhatu{1}$ and $\varhatu{2}$, respectively.
Relative to the unadjusted estimator, both LR and RF deliver point estimates of the same sign and order of magnitude while reducing the standard errors under both variance estimators, which in turn leads to smaller $p$-values.
In particular, RF combined with $\varhatu{2}$ gives the smallest standard error ($0.63\%$) and indicates that the updated fine-ranking formula significantly increases sender-to-receiver interaction intensity at the conventional 5\% level (two-sided $p=0.049$).
The pattern is consistent with the theoretical and simulation results: linear adjustment can improve estimation precision, while a flexible prediction model can provide additional gains when the relationship between covariates and outcomes is nonlinear.
All estimators were implemented in PySpark and deployed on the platform's internal distributed analytical engine as part of the standard experiment-monitoring workflow.

\section{Discussion}
\label{sec:discussion}

This paper develops a design-based framework for causal inference with directed edge-level outcomes under dyadic interference. 
The covariate-adjusted estimators further incorporate pre-treatment edge-level information through three-fold sample splitting and cross-fitting, while the calibration step for non-linear prediction models protects against asymptotic efficiency loss relative to the unadjusted estimator.

The proposed framework is most directly applicable when interference operates primarily through the two endpoints of an edge. 
In settings where higher-order interference may arise, for example through neighbors of neighbors, additional structure is needed to summarize the relevant exposure conditions and control the resulting dependence. 
An extension based on exposure mappings is provided in \Cref{sec.sm:exposure} of the Supplementary Material, but developing inference under this more general interference remains an important direction for future work.

From a practical standpoint, edge-level causal inference requires outcome and covariate information for directed pairs rather than only for individual units. 
This requirement can increase data and computational demands, especially in large-scale platforms, but it also enables a more detailed characterization of how interventions affect interactions between units. 
Researchers should therefore weigh the additional data requirements against the enhanced interpretability afforded by edge-level analysis.

Finally, many practical deployments involve incomplete or noisy covariates and outcomes \citep{zhao2024covariate}. 
Extending the proposed design-based and covariate-adjusted methods to missing-data settings is therefore a promising direction for future research.

\bibliographystyle{agsm}
\bibliography{causal}

\end{document}